\documentclass[aps,pra,amsmath,amssymb,twocolumn,nofootinbib]{revtex4} 

\usepackage{graphicx}
\usepackage{dcolumn}
\usepackage{bm}
\usepackage{bbold}
\usepackage{pslatex}
\usepackage{epsfig}
\usepackage{color}
\usepackage{dsfont}
\usepackage{amsmath}
\usepackage[normalem]{ulem} 

\newcommand{\be}{\begin{equation}}
\newcommand{\ee}{\end{equation}}
\newcommand{\ba}{\begin{eqnarray}}
\newcommand{\ea}{\end{eqnarray}}

\begin{document}
\title[]{Coarse graining a non-Markovian collisional model}

\author{Nadja K. Bernardes$^{1}$\footnote{Corresponding 
author: nadjakb@fisica.ufmg.br}, Andre R. R. Carvalho$^{2}$, C. H. Monken$^{1}$, and Marcelo F. Santos$^{3}$}

\address{$^1$Departamento de F\'isica, Universidade Federal de Minas Gerais, Belo Horizonte, Caixa Postal 702, 30161-970, Brazil}
\address{$^2$Centre for Quantum Computation and Communication Technology, Department of Quantum Sciences, Research School of Physics and Engineering, The Australian National University, Canberra, ACT 0200 Australia}
\address{$^3$Instituto de F\'isica, Universidade Federal do Rio de Janeiro, Caixa Postal 68528, Rio de Janeiro, 21941-972, Brazil}


\date{\today}


\begin{abstract}
The dynamics of systems subjected to noise is called Markovian in the absence of memory effects, i.e. when its immediate future only depends on its present. Time correlations in the noise source may generate non-Markovian effects that, sometimes, can be erased by appropriately coarse graining the time evolution of the system. In general, the coarse graining time $t_\text{CG}$ is taken to be much larger than the correlation time $\tau$ but no direct relation between them is established. Here we analytically obtain a relation between $t_\text{CG}$ and $\tau$ for the dynamics of a qubit subjected to a time correlated environment. Our results can be applied in principle to any distribution of the environmental correlations and can be tested through a collisional model where the qubit sequentially interacts with correlated qutrits.
\end{abstract}

\maketitle

\section{Introduction} 

In open quantum systems two important approximations are made in order to derive a master equation: the Born approximation, which assumes a weak system-environment interaction, and the Markov approximation, which neglects environmental memory effects and is justified when the time-scale characterizing the decay of the environmental correlations is very short in comparison to all the other significant time-scales of the problem \cite{breuer}. It is worth noticing, however, that completely uncorrelated environments are never found in practice. A common procedure to achieve a Markovian master equation then is to coarse grain time, i.e. the evolution of the system is considered in time steps that are much larger than any characteristic time scale of the environment~\cite{lidar, lidar2}, but still short enough when compared to the time scale of changes in the state of the system. In this case, time is discretized in intervals of finite size $t_\text{CG}$ and quantum master equations in the continuous limit, such as in the Lindblad form \cite{lindblad}, have to be considered as an approximation that is good enough if the system is not probed in intervals smaller than $t_\text{CG}$. This restriction is reasonable in many cases and such equations describe a variety of processes like the spontaneous emission of a two-level atom or the damping of an harmonic oscillator \cite{breuer,wiseman}. 

In general, $t_\text{CG}$ is assumed to be much larger than the correlation time of the environment $\tau$, but no specific information about what ``much larger'' actually means is provided. From the experimental point of view, this procedure is partially supported by the fact that any measurement apparatus is limited by a maximum data acquisition rate. But, fundamentally, it means that if one can probe the system fast enough, one should expect to observe non-Markovian effects in its dynamics. This is the basis for the Zeno effect, for example~\cite{zeno}. A natural question, then, regards the limits in time scales within which the non-Markovian dynamics of a given system can be experimentally elucidated. A more detailed relation between $t_\text{CG}$ and $\tau$, however, strongly depends on the specificity of the system and its interaction with the environment and is usually difficult to achieve. It is even more difficult to separate the role played by the correlations in the environment and other non-Markovianity sources such as strong system-environment interactions or the internal dynamics of the environment. In general, all theses factors bunch together when $t_\text{CG}$ is established for a given dynamics.

Recent theoretical~\cite{ziman1,ziman2,giovannetti, tomas, ciccarello, vacchini, budini, Paternostro,nadja} and experimental~\cite{Liu,Liu2,tang,Steve1,Fabio2,chiuri,Xu,Fabio1,nadja2,nadja3} works have demonstrated how to use the so-called collisional models to simulate generic non-unitary dynamics~\cite{ziman1,ziman2} and, specifically, to better understand the emergence of non-Markovian effects~\cite{giovannetti, tomas, ciccarello, vacchini, budini, Paternostro,nadja}. Collisional models are those in which the environment is made of particles that randomly collide with the system, usually one at a time, generating non-unitary dynamics. In particular, in~\cite{nadja}, we have designed a model where the non-Markovianity caused by correlations in the environment can be investigated independently from the other above mentioned factors. 

In this paper, we use the results found in~\cite{nadja} to show how to recover a Markovian dynamics of a qubit by coarse graining its time evolution. In particular, we identify in our model what ``much larger" means by presenting a clear relation between the coarse graining time and the correlation time of the environment. We analytically derive this relation for arbitrary correlation functions of the environment and test our results for two particular types: a step function and an exponential decay of environment correlations. 

In Sec.~\ref{model}, we present our collisional model. The correlation functions are defined in Sec.~\ref{cf}. The dynamics is analyzed in Sec.~\ref{dyn} and the relation between the correlation time and the the coarse grain is established. The limit to the continuous is established in Sec.~\ref{cont_limit}. We conclude in Sec.~\ref{con}. 

\section{The model}\label{model}
In classical physics, the master equation is used to describe the time-evolution of the probability that a system undergoing stochastic evolution is in a particular state at a given time. The quantum mechanical version of this type of evolution for a given physical state $\rho(t)$ is given by the so-called quantum master equation in the Lindblad form
\ba
\frac{d\rho}{dt} = -i[H_s/\hbar,\rho] + \sum_i \mathcal{L}_i(\rho),
\label{Master}
\ea
where $H_s$ is the system Hamiltonian  and $\mathcal{L}_i(\rho) = - \frac{\gamma_i}{2} (\Gamma^\dagger_i \Gamma_i \rho+\rho \Gamma^\dagger_i \Gamma_i - 2 \Gamma_i \rho \Gamma^\dagger_i)$ gives the non-unitary contribution due to the interaction with the environment (responsible for the so-called decoherence of $\rho$). This equation can be rewritten in time steps $dt$ as $\rho(t+dt) = \sum_j W_j \rho(t) W^\dagger_j$ where $W_0 = \mathds{1}-dt(iH_s/\hbar +\sum_i \frac{\gamma_i}{2} \Gamma^\dagger_i \Gamma_i)$, $W_{i \neq 0}=\sqrt{dt \gamma_i}\Gamma_i$, $\gamma_i dt \ll 1$ and terms of $O(dt^2)$ are discarded. In order to guarantee that $\rho$ is a quantum state at any given time, $\gamma_i$ has to be a positive rate for all $i$, and $\sum_j W^\dagger_j W_j = \mathds{1}$. In this way, the trace and positivity of $\rho$ are always preserved. Note that if we assume that $H_s$ is equal to zero and$\rho$ describes the state of a qubit under a random unitary evolution with $\Gamma_i \equiv \sigma_i$ Pauli-like matrices ($\Gamma_i^\dagger = \Gamma_i$ and $\Gamma_i^2 = 1$), then Eq~(\ref{Master}) can be rewritten as the Completely Positive and Trace Preserving (CPTP) map 
\ba\label{rhof}
\Lambda_{n+1,n}:\rho_{n}\mapsto\rho_{(n+1)}&=&\Lambda_{n+1,n}(\rho_{n})\\
&=&(1-\sum_{i=1,2} \epsilon_i)\rho_{n}+\sum_{i=1,2} \epsilon_i \sigma_i\rho_{n}\sigma_i.\nonumber
\ea
where $\epsilon_i \equiv \gamma_i dt$ gives the probability of a respective change $\sigma_i$ in the system in an arbitrary time interval $\{ndt,(n+1)dt\}$. We will also consider only two channels, $\sigma_1$ and $\sigma_2$ with no loss of generality, as previously shown in~\cite{nadja}. This is the scenario to be explored from now on.

This kind of time evolution, specially the discretized time version described in Eq.(\ref{rhof}), has been associated to collisional models \cite{Plenio3}. In this case, each time step $n \rightarrow n+1$ corresponds to a collision with the reservoir and the Markovian hypothesis is that the map after $m$ collisions, that takes the system from $n$ to $n+m$, can be concatenated as the product of $m$ intermediate CPTP maps: $\Lambda_{n+m,n}=\Lambda_{n+m,n+m-1}\Lambda_{n+m-1,n+m-2}...\Lambda_{n+1,n}$, for arbitrary $n$ and $m$. However, as shown in~\cite{nadja}, correlations in the environmental state may lead to time dependent rates when the probabilities $\epsilon_i$ depend on $n$. That happens even if the transformations undergone by the system are equal ($\sigma_1 = \sigma_2$). If, on the other hand, $\sigma_1 \neq \sigma_2$, environmental correlations also produce ``artificial'' decoherence channels (extra $W$ operators) that do not derive directly from the collisions but from their correlation, and, more important, that may be associated to negative rates $\gamma_i <0$~\cite{nadja}. In this case, Eq.~(\ref{rhof}) does not describe correctly the physical process ($\rho_{n+1}$ cannot be described as CP map applied to $\rho_{n}$) and the dynamics becomes non-Markovian, i.e. the state of the qubit after a given collision becomes dependent not only on its state immediately before it but also on its state at earlier times. 

In~\cite{nadja} we analyzed the effects of correlations in just two consecutive collisions because we were interested in understanding the minimum effects that environmental correlations could produce. Physically, this situation corresponds to a very short correlation pulse in an otherwise uncorrelated environment. In the present work we are interested in a more realistic scenario for the environment so that we can address the opposite question: given that most physical environments are always somewhat correlated and that these correlations may generate non-Markovian evolutions, we investigate the minimum coarse graining in time that one needs to define in this model so that any observation on the system concludes for a Markovian time evolution in the Lindblad form.  Note that to coarse grain time in our model corresponds to allow for a certain number of collisions to occur in between any consecutive observations of the system. In terms of the quantities already defined, it means to establish a minimum interval of collisions $n_\text{CG}$ so that $\rho_{(n+n_\text{CG})}=\sum_j W'_j \rho_{n} W'^{\dagger}_j$, for any ``time'' $n$, where the set $\{W'_j\}$ shares the same general properties of the previous $\{W_j\}$ set even though the operators and rates will not necessarily be the same.

\section{Correlation functions}\label{cf}

In order to better understand how environment correlation functions are introduced in our model, it is useful to visualize its uncorrelated version in terms of quantum trajectories. If the qubit is initially prepared in a pure state $|\Psi_0\rangle$, then a quantum trajectory $Q$ that takes the system from its initial state to its state $|\Psi_n\rangle$ after $n$ collisions ($T=ndt$) corresponds to a sequence $\{|\Psi_0\rangle,|\Psi_1\rangle,|\Psi_2\rangle,...,|\Psi_n\rangle\}$, where $|\Psi_k\rangle = U^{(k)}|\Psi_{k-1}\rangle$ and $U^{(k)}$ is drawn from the set $\{\sigma_0,\sigma_1,\sigma_2\}$ ($\sigma_0=\mathds{1}$) with corresponding probabilities $\{p_k(0),p_k(1),p_k(2)\}$ given by $\{ 1-(\epsilon_1+\epsilon_2),\epsilon_1,\epsilon_2\}$. A trajectory $Q$ from $0$ to $n$ is then defined by the sequence $Q|\Psi\rangle= \Pi_{k=1}^{n} U^{(k)}|\Psi_0\rangle$ which happens with probability $P_Q=\Pi_{k=1}^n p_{k}$, where $p_k$ is the probability associated to each particular realization of $U^{(k)}$. The state of the system at $n$ is given by averaging over all possible trajectories $\rho_{n} = \sum_Q Q\rho_{0} Q^\dagger$ and, more important, Eq.~(\ref{rhof}) holds for any ``time'' $n$ when $\epsilon_i \ll 1$.

Correlations in the reservoir are introduced by modifying the condition just described, i.e. the joint probabilities $p_{kk'}(i,j)$ of two (or more) jumps of the same type become more likely to happen then for independent, random events: $p_{kk'}(i,i) \neq \epsilon_i^2$, where $p_{kk'}(i,j)$ is the joint probability that $i$-th jump occurred in the $k$-th collision and the $j$-th jump occurred in the $k'$-th collision. In this work, we analyze two models of correlation functions for the environment: a step function and an exponential time decay. In both cases the correlation time of the reservoir is well defined as we explain later. For the step function, we choose a certain number of collisions for which once the first jump is of a given type, say $\sigma_1$, then any other jump in the same interval must be of the same type ($p_{kk'}(1,2)=0$ within the interval). In this scenario, this interval is equivalent to the correlation time of the reservoir from now on denoted by $n_\text{cor}$. This choice naturally modifies the probabilities in each step and clearly introduces memory in the dynamics. First note that since $\epsilon_i \ll 1$ the most likely operation in each time step still is the no-jump $W_0$ associated to applying identity to the state of the system. Second, note that the counting of the correlated time interval always starts at the first jump.
From then on and for the duration of $n_\text{cor}$, for that particular trajectory any other possible jump has to be the same. 

The exponential decay model follows the same logic, the difference being that now once a jump happens, the probability for the same type of jump to happen in the near future decays exponentially as a function of time (number of collisions), with the joint probability given by: $p_{kk'}(i,i)\propto \epsilon^2(1+e^{\frac{-(k'-k)}{n_\text{cor}}})$, where $i=1,2$. In this case, $1/n_\text{cor}$ defines the decay rate of the reservoir correlations. In both cases, the effects of the reservoir correlations in the dynamics of the system are analyzed through calculating the map $\Lambda_{(n+m),n}$ that takes $\rho_{n}$ into $\rho_{(n+m)}$, for arbitrary $n$ and number of collisions $m$, and testing the eventual deviations from the map shown in Eq.~(\ref{rhof}).

\begin{figure}[h]
(a)		
\includegraphics[width=0.48\textwidth]{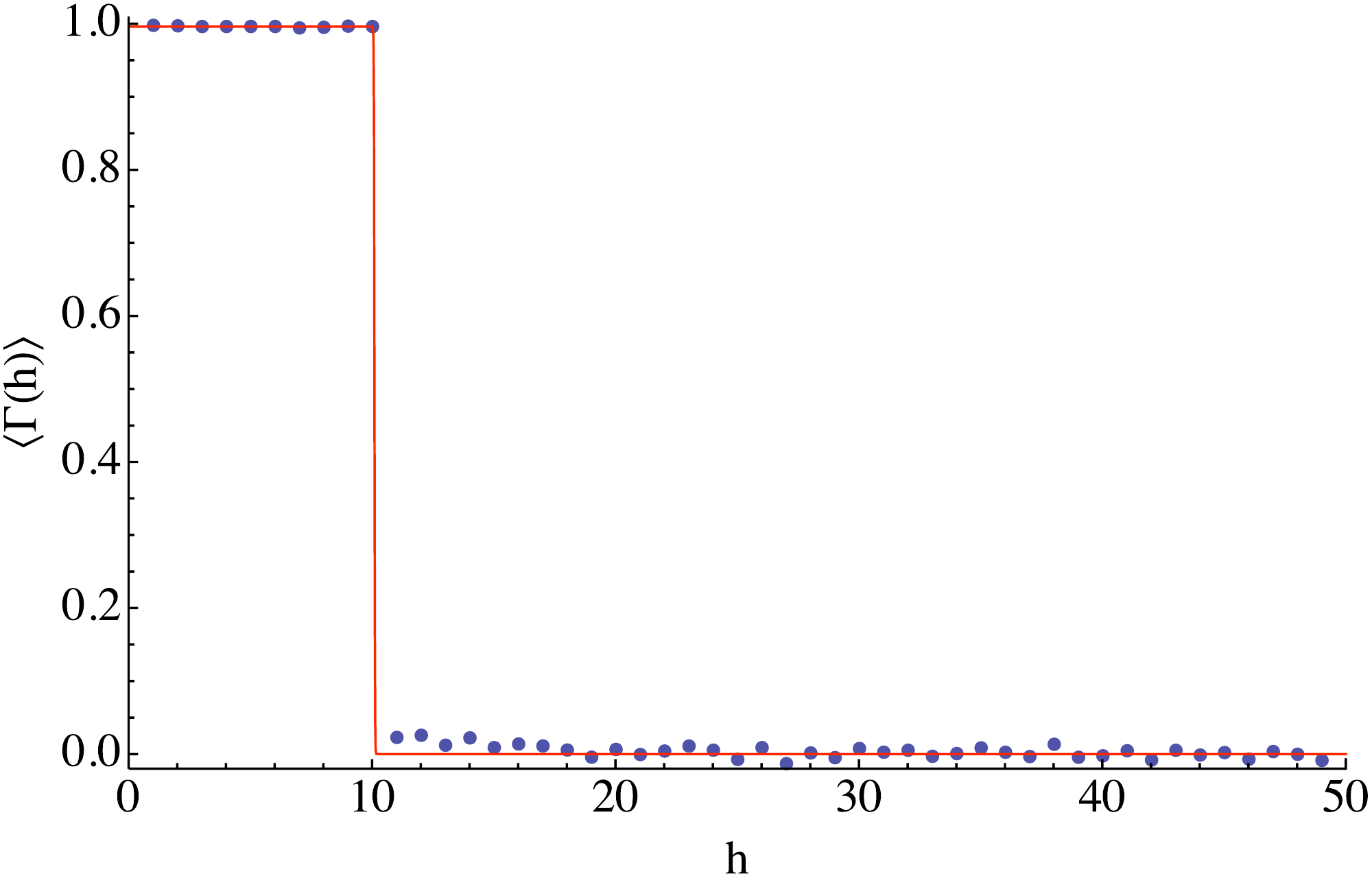}
(b)
\includegraphics[width=0.48\textwidth]{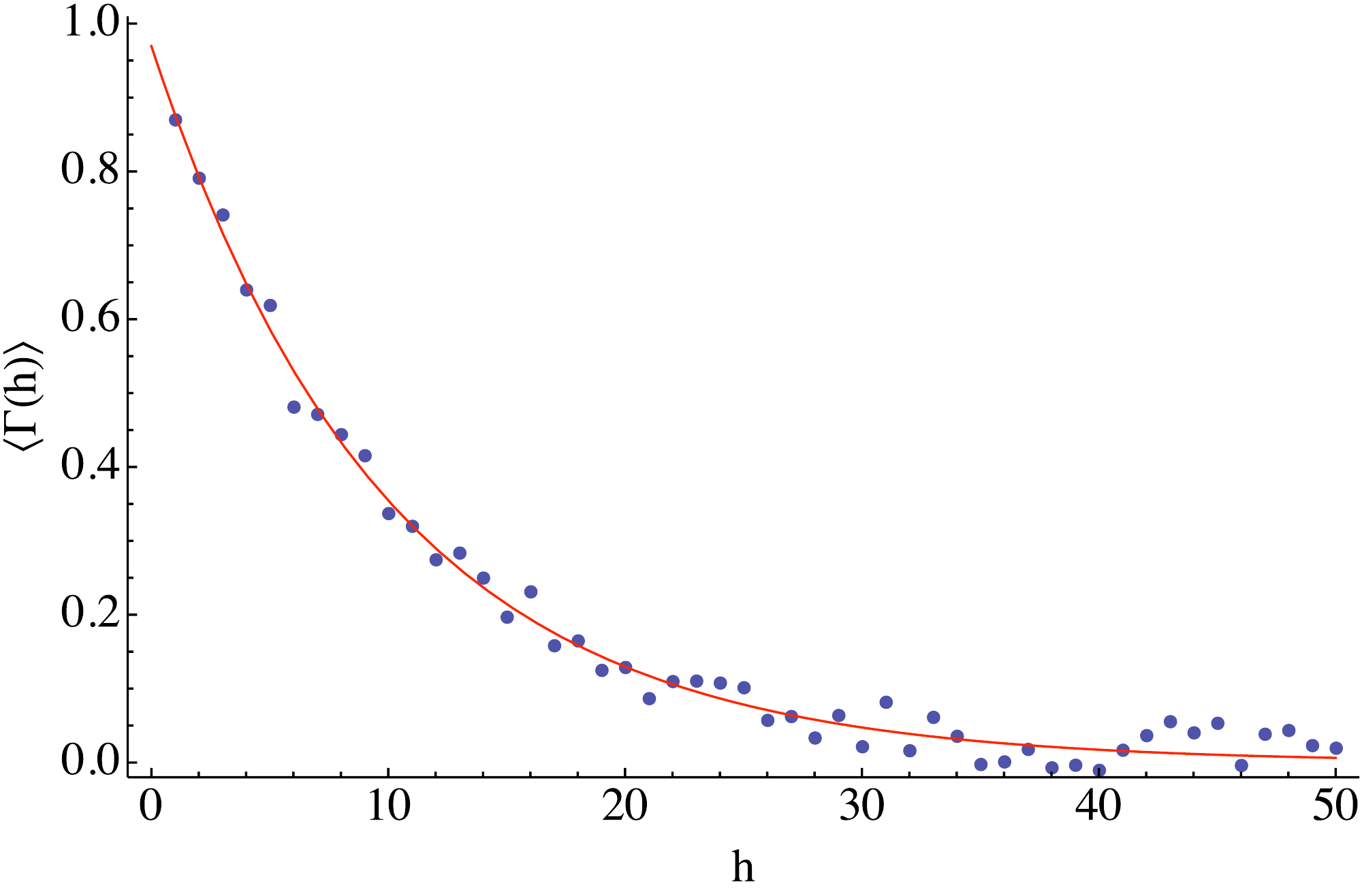}
	\caption{Plot of the average correlation function $\left<\Gamma(h)\right>$ for $10^7$ trajectories and the total number of collisions of $n=100$ for (a) the step function and (b) the exponential decay. The red line represents a fit, where the estimated correlation length is $\bar{n}_\text{cor}=10.009$ in (a) and $\bar{n}_\text{cor}=10.28$ in (b) in agreement with the expected value of $n_\text{cor}=10$. }
	\label{figcorr}
\end{figure}

Following the results of~\cite{nadja}, it suffices to investigate the configuration in which the probabilities of the two different channels $\sigma_1$ and $\sigma_2$ are equal ($\epsilon_1=\epsilon_2=\epsilon$), with $\sigma_1 = \sigma_x$ and $\sigma_2=\sigma_z$ (corresponding to bit flip and dephasing). This type of decoherence is enough to encompass most of the interesting effects the model can produce; any other configuration implies in less non-Markovianity per time step. To illustrate the functions and confirm that the numerical simulation is in agreement with the theoretical model, we plotted in Fig.~\ref{figcorr} the average correlation function $\left<\Gamma(h)\right>$ for different trajectories, where $\Gamma(h)=\frac{\sum_{i=1}^{n-h}x_ix_{i+h}}{\sum_{i=1}^{n-h}|x_ix_{i+h}|}$ with $x_i=1$(-1) if the collision is $\sigma_x$($\sigma_z$). 

\section{Dynamics analysis}\label{dyn}

When $\sigma_1 = \sigma_x$, $\sigma_2=\sigma_z$ and $\epsilon_1 = \epsilon_2= \epsilon$, correlations in the reservoir end up generating non-Markovian evolutions that can be evidenced by the non-complete-positivity divisibility of the maps~\cite{nadja}, i.e. CP maps for the overall evolution over $m$ collisions that may not be decomposable into the product of CP maps for the intermediate evolutions, $\Lambda_{m+i,i} \neq \Lambda_{m+i,n+i}\Lambda_{n+i,i}$ ($n<m$). In this case, finding the coarse graining time means finding the minimum $n$ beyond which the decomposition into CP maps is always possible. In order to find $n_\text{CG}$ for each model of correlation in the reservoir, we test the maps $\Lambda_{m+i,i}$ that take $\rho_{i}$ into $\rho_{i+m}$ for different intervals $m$. Testing the complete positivity of a map is equivalent to testing whether the Choi matrix (also known as the dynamical matrix) associated to this map is positive semidefinite, i.e. it has nonnegative eigenvalues \cite{choi,bengtsson}.

\begin{figure}[h]
(a)		
\includegraphics[width=0.48\textwidth]{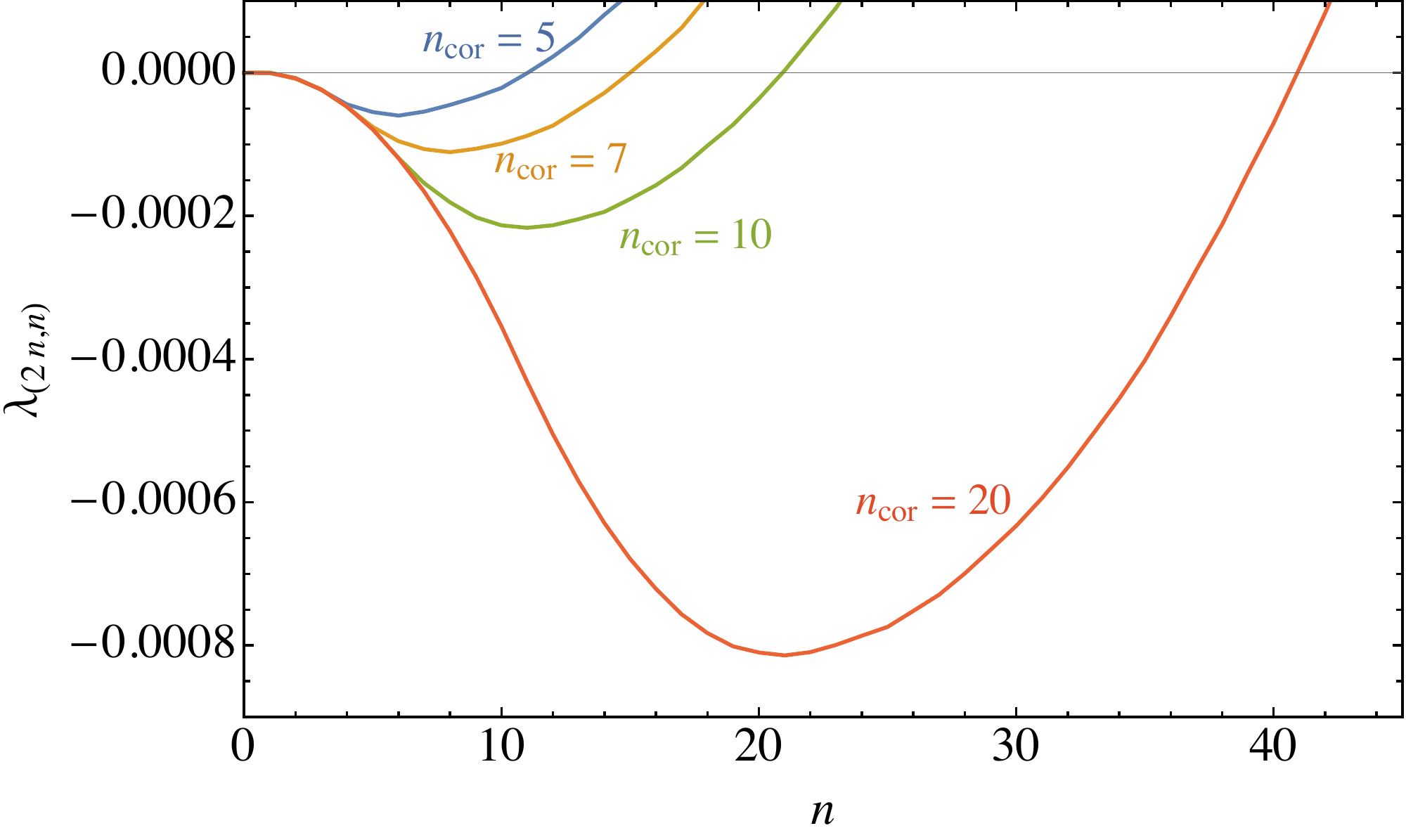}
(b)
\includegraphics[width=0.48\textwidth]{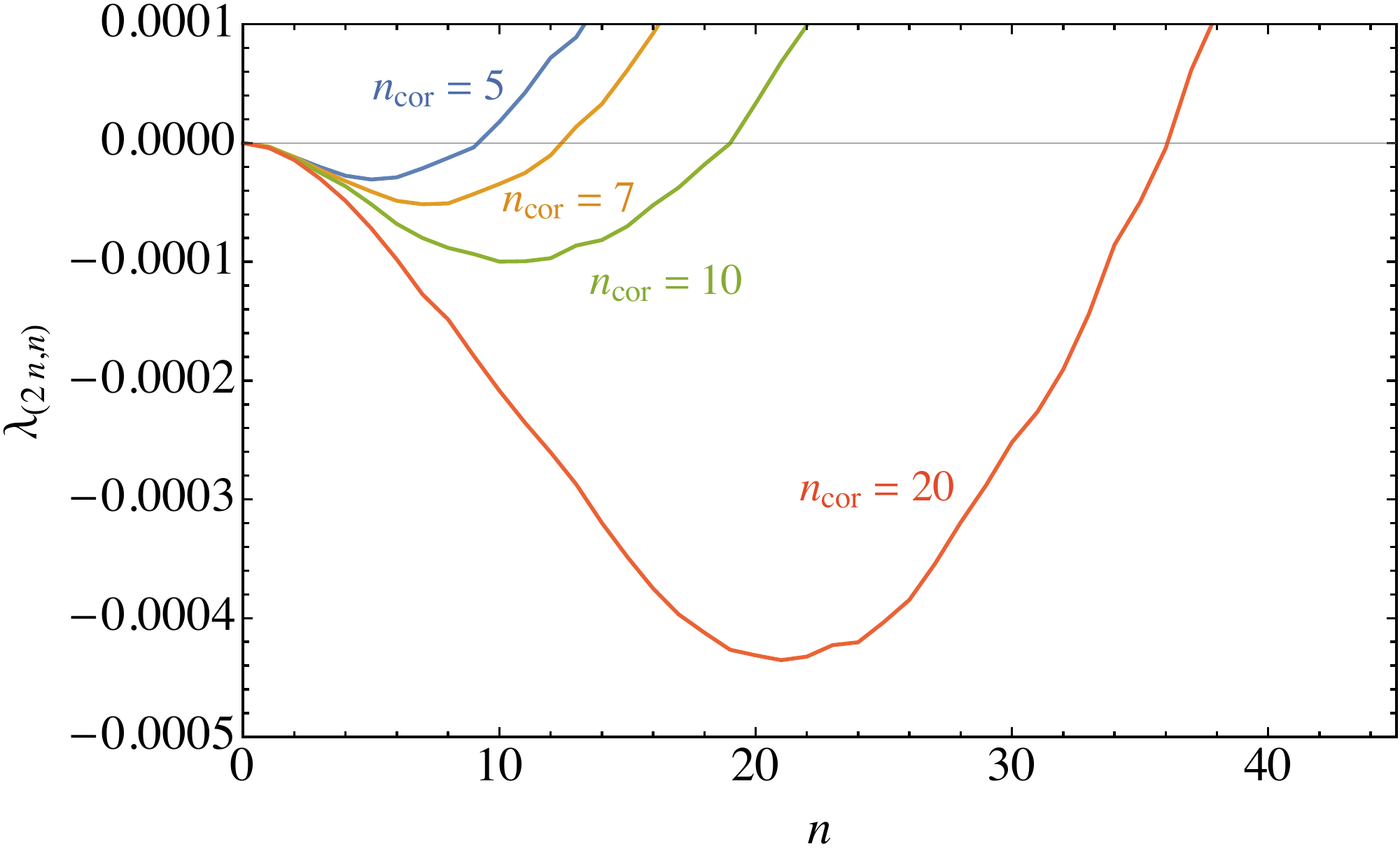}
	\caption{Plot of the smallest eigenvalue $\lambda_{2n,n}$ of the dynamical matrix of the map $\Lambda_{2n,n}$ versus the number of collisions $n$, for $\epsilon=0.001$, $10^8$ trajectories and for (a) the step function and (b) the exponential decay function. Different lines represent different correlation lengths as it is shown in the plot. The lines are just for illustrative reasons, since the number of collisions is a discretized quantity.}
	\label{lambdadegrau}
\end{figure}

We will analyze the case where $i=0$ and $m=2n$. In Fig.~\ref{lambdadegrau} the smallest eigenvalue $\lambda_{2n,n}$ of the dynamical matrix corresponding to the map $\Lambda_{2n,n}$ is plotted versus the number of collisions $n$ for the step function and for the exponential decay function. If this eigenvalue is positive, this means that $\Lambda_{2n,0}$ is divisible in $\Lambda_{2n,0}=\Lambda_{2n,n}\Lambda_{n,0}$.  We can identify here what are the points where the curves cross the $\lambda=0$ axis, i.e. it assumes positive values again, and consequently identify a coarse graining length $n_\text{CG}$ as a function of $n_\text{cor}$. 

\begin{figure}[t!]
	\includegraphics[width=0.48\textwidth]{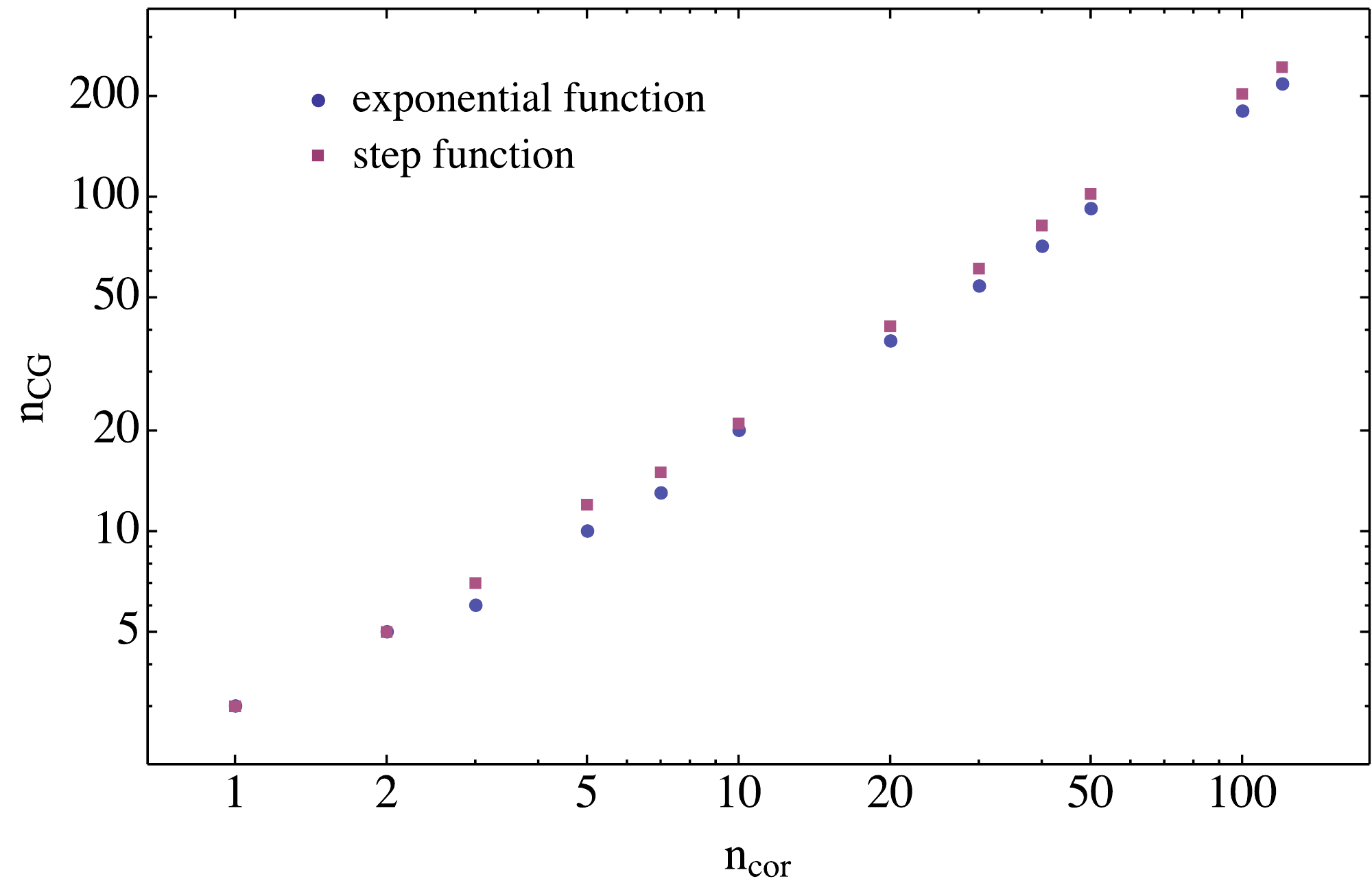}
	\caption{Plot of the coarse grain $n_\text{CG}$ versus the correlation length $n_\text{cor}$ for the exponential function (dots) and the step function (squares) for $\epsilon=0.001$ and $10^8$ trajectories. }
	\label{ncorxncg}
\end{figure}

By identifying a coarse grain $n_\text{CG}$ for the step and exponential correlation function, we are able to show the dependence of the coarse grain with the correlation length as presented in Fig.~\ref{ncorxncg}.  It is possible to see that the size of the coarse grain is increasing almost linearly with the correlation length by a factor of approximately $2$.

This result can be explained by analyzing the maps of the resulting dynamics. The map from one ``time" $i$ to $j$ is given by $\Lambda_{j,i}(\rho(i))=\sum_{\alpha=0,x,y,z}w_{\alpha}(j,i)\sigma_{\alpha}\rho(i)\sigma_{\alpha}$. Note here that although in one collision, the system may undergo only a $\sigma_x$ or a $\sigma_z$ operation, with more collisions a $\sigma_y$ operation may result from the combination of $\sigma_x$ and $\sigma_z$. If the environment is uncorrelated, and to second order in $\epsilon$, this somewhat artificial $\sigma_y$ channel, that is created by bunching collisions, is proportional to the chance $\epsilon^2$ of having two collisions of different types in that particular bunch (always remembering that in each time step, the most probable collision is of the $\sigma_0=\mathds{1}$ type). Truncating in $O(\epsilon^2)$ is justifiable for $\epsilon \ll 1$ (we will comment more on this later). In fact, for the uncorrelated case, the probability of such $\sigma_y$ channel is the same as that of an extra $\sigma_0$ contribution to the dynamics, since the chance of similar or opposite collisions of type $\sigma_x$ and $\sigma_z$ is the same. However, when correlations in the environment are introduced, these probabilities become unbalanced and the extra $\sigma_0$ channel becomes more probable then the $\sigma_y$ one. This unbalance is the origin of the non-Markovian effect and the ``lack'' of some portion of the $\sigma_y$ channel reflects on a negative weight $w_y$ in any attempt to force the overall map corresponding to that bunch to be a concatenation of maps, each one representing a single collision. However, due to the finitude of $n_\text{cor}$, if one takes a large enough interval $n_\text{CG} > n_\text{cor}$, the uncorrelated events in this interval compensate for the memory effects brought by the correlated collisions and the dynamics coarse grained in $n_\text{CG}$ intervals become Markovian. Therefore, in order to find $n_\text{CG}$, one needs to check the minimum time interval beyond which all the weights, including that of $\sigma_y$, become nonnegative. In this model, since the map is invertible, this is equivalent to check if $\Lambda_{2n,n}=\Lambda_{2n,0}\Lambda_{n,0}^{-1}$ is a CPTP map, which already imposes constraints over the weights $w_{\alpha}(2n,n)$, i.e. $w_{\alpha}(2n,n)\ge0$. This is exactly what was numerically done and presented in Fig.3.

However, for small enough $\epsilon$ (in our calculations $\epsilon=0.001$), and as long as the bunching of $M$ collisions is not too large, i.e. $M\epsilon$ is still much smaller than one, the chance of three or more collisions of type $\sigma_x$ or $\sigma_z$ within this bunch is very low and one can assume at most two collisions of these types with no significant loss of generality (in our numerics, out of $10^8$ generated trajectories, less than 5\% would present three or more of such collisions). This allows us to analytically calculate a very good approximation for the weights $w_{\alpha}$, one that becomes even better in the continuous limit of $\epsilon \rightarrow 0$. First, let us define the weights as:

\ba
\label{coefs}
w(n,0)&=&\frac{1}{2}\binom{n}{1}p(1-p)^{n-1},\\
\label{coefsb}
w_{y}(n,0)&=&\frac{1}{2}\binom{n}{2}p^2(1-p)^{n-2}\nonumber\\
&&\times\left[1 - \sum_{k=1}^{n-1}f(k, n_{\text{cor}}) G(k, n, p)\right],
\ea
where $p=2\epsilon$ is the probability that in one collision the system will undergo a $\sigma_x$ or $\sigma_z$ rotation for an uncorrelated environment and we take $w_x=w_z=w$ since both channels are symmetric in our model. $f(k, n_\text{cor})$ is the correlation function of the environment and for the examples studied in this paper, step or exponential decaying function, it assumes the respective forms
\ba
\label{function}
f_\text{step}(k,  n_{\text{cor}})&=&\left\{ \begin{array}{c}
			1\quad \text{if}\quad k \leq  n_{\text{cor}},\\
			0 \quad \text{otherwise},
\end{array}
\right.\\
f_\text{exp}(k, n_{\text{cor}})&=&\text{e}^{-\frac{k}{ n_{\text{cor}}}}.
\ea
Note, however, that at this point the model is generic enough to consider any function $f(k, n_\text{cor})$.

The function $G(k, n, p)$ is related to the probability that two collisions of the type $\sigma_x$ or $\sigma_z$ occur separated by $k$ collisions in an interval of $n$ collisions and it is given by
\be
\label{G}
G(k, n, p)= \frac{p^2(n - k) (1 - p)^{k - 1} }{(1 - p)^n + 
  n p - 1}.
\ee

The intermediate $\Lambda_{2n,n}$ map will be given $\Lambda_{2n,n}(\rho(n))=\sum_{\alpha=0,x,y,z}w_{\alpha}(2n,n)\sigma_{\alpha}\rho(n)\sigma_{\alpha}$, where the coefficients are given in terms of the coefficients from Eq.~(\ref{coefs},\ref{coefsb}) as

\ba
\label{coef2}
w_{0}(2n,n)&=&1-2w(2n,n)-w_{y}(2n,n),\\
\label{coef2s}
w(2n,n)&=&\frac{w(2n,0)-w(n,0)}{1-4w(n,0)},\\
\label{coef2sb}
w_{y}(2n,n)&=&\frac{1}{4}\left[1 +\frac{1-4w(2n,0)}{1-4w(n,0)}\right.\\
&&-\left.\frac{2-4w(2n,0)-4w_y(2n,0)}{1-2w(n,0)-2w_y(n,0)}\right],\nonumber
\ea
where, once again, we have considered $w_{x}(j,i)=w_{z}(j,i)=w(j,i)$ due to the symmetry of $\sigma_x$ and $\sigma_z$ channels.

For the reasons previously discussed regarding the origin of the $\sigma_y$ channel or a simple inspection of Eqs.~(\ref{coef2}-\ref{coef2sb}), $w_y$ is the only weight that can assume negative values. Thus, the map $\Lambda_{2n,0}$ is divisible, if $w_{y}(2n,n)\geq0$. Note that this corresponds exactly to verifying when the eigenvalues of the dynamical matrix are all positive. In fact, it is possible to show that $w_y=2\lambda_y$ and, for small $\epsilon$, the inequality $\lambda_y \geq 0$ recovers the result observed in the numerical simulations for the coarse grain time $n_\text{CG}$. For example, for the step function, if one inserts Eqs.~(\ref{function},\ref{G}) into Eqs.~(\ref{coef2}-\ref{coef2sb}) and then expands the latters keeping terms up to $O(\epsilon ^2)$, the corresponding inequality for $w_y(2n,n)$ simplifies to 
\ba
n(n-1-2n_\text{cor})>0,
\ea
which has a solution for $n>2 n_{\text{cor}}+1$. A similar result can be found for the exponential function. In this way, we confirmed the previous result encountered in the numerical simulation presented in Fig.~\ref{ncorxncg}.

\section{Continuous limit}\label{cont_limit}
So far, we have considered a discretized model of time steps of finite size proportional to $\epsilon$. Note, however, that in our model $\epsilon$ can be taken as small as desired, which means that we can take the model to the continuous limit. In order to do this, we make the following change of variables in Eqs.~(\ref{function}-\ref{G}): $n=T/\Delta$, $k=t/\Delta$, $n_{\text{cor}}=\tau/\Delta$, and $\epsilon=\gamma \Delta$. The functions then change to:
\ba
f_{\text{step}}(t,\tau)&=&\left\{ \begin{array}{c}
			1\quad \text{if}\quad t \leq  \tau,\\
			0 \quad \text{otherwise},
\end{array}
\right.\\
f_\text{exp}(t,\tau)&=&\text{e}^{-\frac{t}{\tau}},\\
G\left(t, T, \gamma \right)&=&\frac{4\gamma^2 (T-t)(1-2\gamma\Delta)^{\frac{t}{\Delta}-1}\Delta}{(1-2\gamma\Delta)^{\frac{T}{\Delta}}+2\gamma T-1}.
\ea

In this way, with fixed $\gamma$, when $\Delta \rightarrow 0$, it follows that $n \rightarrow \infty$ and $\epsilon \rightarrow 0$, but we guarantee that $n\epsilon=\gamma T$. As in the Riemman-Steltjes sum, the limit of the sum will be given by
\be
\label{Riemman}
\lim_{\Delta\rightarrow 0}\sum_{t=\Delta}^{T-\Delta}f\left(\frac{t}{\Delta},  \frac{\tau}{\Delta}\right)G\left(\frac{t}{\Delta}, \frac{T}{\Delta}, \gamma \Delta\right)=\int_{0}^{T}f\left(t,\tau \right)g(t,T,\gamma) dt,
\ee
where
\be
g\left(t,T,\gamma\right)=\frac{4\gamma^2 (T-t)\text{e}^{-2\gamma t}}{\text{e}^{-2\gamma T}+2\gamma T-1}.
\ee
Eq.~(\ref{Riemman}) does not depend on the particular form of the correlation function and, therefore, this collisional model (determined by the value of $\epsilon$) indeed stroboscopically simulates a continuous-time quantum evolution for a variety of different reservoir correlation functions.

\section{Conclusions}\label{con}
In this work we have used a previously designed collisional model for the non-Markovian evolution of a qubit to show that by coarse graining its dynamics in time, i.e. defining minimal time intervals to probe the system, a Markovian evolution can be recovered. We have also shown that our model allows one to properly identify a relation between the correlation time of the environment causing the non-unitary (and non-Markovian) evolution of the qubit and the coarse graining time that restores its Markovianity. We test the model for two types of correlation, a step function and an exponential decay and we show that, surprisingly, depending on the correlation function, the ratio between the coarse graining time and the correlation time, which is usually taken to be much larger than one, can be as small as two. The model is general enough to encompass different correlation functions, hence a broad variety of environments, and we have derived an approximate analytical solution that fits well our numerical results, including a limit to the continuous. Finally, the results presented here could be also useful to identify how broad the spectral response of a detector should be in order that memory effects caused by the environment can be neglected. 

\subsection*{Acknowledgments}
 N.K.B, C.H.M. and M.F.S. would like to thank the support from the Brazilian agencies CNPq and CAPES. M.F.S. would like to thank the support of FAPEMIG, project PPM IV. This work is part of the INCT-IQ from CNPq and also of the Australian Research Council Centre of Excellence for Quantum Computation and Communication Technology (project number CE110001027).

\end{document}